\documentclass[prd,showpacs,preprintnumbers,twocolumn,amsmath,nofootinbib,amssymb,floatfix]{revtex4}
\usepackage{graphicx,color,dcolumn,booktabs,bm}
\usepackage{epstopdf}
\usepackage{longtable,lscape}
\usepackage{txfonts}
\usepackage{overpic}
\usepackage{float}
\usepackage{amssymb}
\usepackage{amsmath}
\usepackage{epstopdf}
\usepackage{indentfirst}
\usepackage{feynmf}  
\usepackage{slashed}  
\usepackage{cases}
\usepackage{ulem}
\usepackage{color}
\usepackage{float}
\usepackage{multirow}
\usepackage{tikz}
\usepackage{epstopdf}
\usepackage{epsfig,dsfont,amssymb,amsmath,amsfonts,amsbsy,mathrsfs}
\usepackage{multirow}
\usepackage{graphicx}  
\usepackage{float}  
\usepackage{subfigure}  
\usepackage{array}
\usepackage{microtype}
\usepackage{pdfpages} 

\usepackage[colorlinks, citecolor=blue,anchorcolor=blue,menucolor=blue, linkcolor=blue,filecolor=blue,runcolor=blue,urlcolor=blue,frenchlinks=true]{hyperref}
\makeatletter
\@addtoreset{equation}{section}
\makeatother

\newcommand{\nc}{\newcommand}
\nc{\tj}[1]{\textcolor{red}{Tianjin: #1}}
\newcommand{\red}[1]{{\color{black} {#1}}}

\newcommand{\purple}[1]{{\color{black} {#1}}}
\usepackage{soul}
\allowdisplaybreaks
\begin{document}
\title{High spin kaons}
\author{Ya-Rong Wang$^{1,2}$}
\author{Hao Chen$^{3}$}
\author{Xiao-Hai Liu$^{1}$}\email{xiaohai.liu@tju.edu.cn}
\author{Cheng-Qun Pang$^{4}$}\email{xuehua45@163.com}
\affiliation{ $^1$Center for Joint Quantum Studies and Department of Physics, School of Science, Tianjin University, Tianjin 300350, China,\\
$^2$Center for Theoretical Physics, School of Physics and Optoelectronic Engineering, Hainan University, Haikou 570228, China,\\
$^3$College of Physics and Electronic Information Engineering, Qinghai Normal University, Xining 810000, China,\\
$^4$School of Physics and Optoelectronic Engineering, Ludong University, Yantai 264000, China\\
}

\date{\today}

\begin{abstract}
The COMPASS Collaboration recently reported the observation of strange-meson spectra in the reaction $K^- + p \to K^- \pi^- \pi^+ + p$ and found $K_3$ and $K_4$ states, with masses of $2119 \pm 13 ^{+45}_{-12}$ MeV and $2210 \pm 40 ^{+80}_{-30}$ MeV, respectively. This discovery has significantly renewed interest, prompting a detailed and systematic study of high-spin kaons. In this work, we analyze the mass spectrum and the Okubo-Zweig-Iizuka-allowed two-body strong decay properties of high-spin kaons having $J^P=3^{\pm}, 4^{\pm}$, and $5^{\pm}$ within the framework of the modified Godfrey-Isgur model and the $^3P_0$ model. Moreover, we identify critical decay channels, which may serve as useful guidance for future experimental studies.

\end{abstract}
\maketitle

\section{introduction}\label{1} 
The complex nature of strong interactions results in a rich spectrum  within quantum chromodynamics (QCD), encompassing bound states and resonances. Among these, the properties of strange mesons remain relatively poorly understood. Therefore, dedicated theoretical investigations are essential--not only to clarify the structure of strange mesons, but also to bridge the gap between the spectra of light mesons and heavy-light mesons. Such efforts are crucial for delineating the excitation spectrum of strange mesons.

Recently, the COMPASS Collaboration at CERN identified $K_3(2120)$ and $K_4(2210)$ states in the scattering reaction $K^- + p \to K^- \pi^- \pi^+ + p$ using the COMPASS spectrometer \cite{COMPASS:2025wkw}. The masses of these two states are determined to be $2119 \pm 13 ^{+45}_{-12}$ MeV and $2210 \pm 40 ^{+80}_{-30}$ MeV, respectively. Concurrently, the widths of these states were measured as $270 \pm 30 ^{+40}_{-30}$ MeV for $K_3(2120)$ and $250 \pm 70 ^{+50}_{-70}$ MeV for $K_4(2210)$.  
This discovery motivates a systematically examination of the high-spin properties of kaons. 

Extensive research has been conducted on low-lying strange mesons over the past decades. 
In 1985, Godfrey and Isgur calculated the mass spectrum of strange mesons using the relativized quark model, which effectively describes the low-lying spectrum reasonably \cite{Godfrey:1985xj}.  
Ebert \red{\textit{et al}}. investigated the mass spectrum and Regge trajectories of kaons within the context of a relativistic quark model in 2009  \cite{Ebert:2009ub}. Significantly, their study  predicted the existence of numerous excited states of kaons \cite{Ebert:2009ub}. 
In 2015, researchers from the Lanzhou group explored the pseudoscalar states $\pi_2$, $\eta_2$, and $K_2$ via the Okubo-Zweig-Iizuka (OZI)-allowed two-body strong decays \cite{Wang:2014sea}. 
After two years, the Lanzhou group examined the mass spectrum of the kaon family using the modified Godfrey-Isgur (MGI) model \cite{Pang:2017dlw}. 
In 2022, Feng \red{\textit{et al}}. studied the mass spectrum of $1^{1}D_2$ and $1^{3}D_2$ meson nonets within the framework of meson mass matrices and Regge phenomenology, highlighting the mixing of the $K_2(1820)$ state with $K_2(1770)$ state \cite{Feng:2022jtg}.
In the same year, Vanamali Shastry presented the results of a theoretical study involving the lightest $1^{-+}$ hybrid kaons predicted by nonrelativistic quark models and observed experimentally \cite{Shastry:2022upd}.
In 2023, \red{a} systematic investigation of excited kaons was carried out, yielding valuable insights into the properties of excited strange mesons \cite{Oudichhya:2023lva}.
Nieto \red{\textit{et al}}. revisited the kaon spectrum by employing a constituent quark model that incorporates the quark-antiquark interaction, which includes the \red{nonperturbative} phenomena of dynamical chiral symmetry breaking and color confinement, as well as the perturbative one-gluon exchange in the same year \cite{Taboada-Nieto:2022igy}. 
In 2024, Li \red{\textit{et al}}. studied the strong decay behaviors of excited axial-vector strange mesons within the quark pair creation model. Their results indicated that the $K_1(1793)/K_1(1861)$ can be regarded as the same $K_1(2P)$ state, and the $K_1(1911)$ is assigned as the $K_1^\prime(2P)$ state \cite{Li:2024hrf}. 
All these studies pertain predominantly to low-lying strange mesons, and a few of \purple{researches} conducted on high-spin kaons. In addition, these studies are exclusively focused on spectroscopy, nevertheless, the study of decay behaviors is rare.

Up to the present, the Particle Data Group (PDG) has listed 28 strange mesons that have been identified within the energy range $[0.5, 3.1]$ GeV. 
Among the 28 states, only 17 kaons are well established whereas the remaining 11 states still require further confirmation. Notably, there are five high spin kaons among these 28 states \cite{ParticleDataGroup:2024cfk}.   
\red{What is} more, the complexity of strange mesons makes it challenging to explain all experimental data simultaneously. 
Broadly speaking, high spin kaons have not yet been studied systematically and comprehensively.

The MGI model performs well for the meson spectra especially high excited states \cite{Feng:2022hwq, Wang:2021abg, Wang:2024yvo}. The screening potential effect has a bigger influence on the high spin kaons. Therefore, in this work, we investigate the mass spectrum of $J^P=3^{\pm}, 4^{\pm}$, and $5^{\pm}$ high spin kaons using \purple{the MGI model. } 
In the meantime, we examine the mass spectrum by Regge trajectory. Base on the analysis of \purple{the mass} spectrum, we study strong decay behaviors of these high spin kaons via the $^3P_0$ model, which is widely applied to the OZI-allowed two-body strong decays of mesons \cite{vanBeveren:1982qb, Titov:1995si, Ackleh:1996yt, Blundell:1996as, Bonnaz:2001aj, Zhou:2004mw, Lu:2006ry, Zhang:2006yj, Luo:2009wu, Sun:2009tg, Liu:2009fe, Sun:2010pg, Rijken:2010zza, Ye:2012gu, Wang:2012wa}.
Additionally, we will identify several key decay channels that offer essential guidance for future experimental investigations.

This paper is structured as follows. After the introduction, we \purple{provide} a concise overview of the Regge trajectory, the MGI model, and the $^3P_0$ model in \red{Sec.} \ref{2}. In \red{Sec.} \ref{3}, the numerical results are presented.      
The paper \purple{ends} with a concise summary in \red{Sec.} \ref{4}.

\section{Models employed in this work} \label{2}

\subsection{Regge trajectory}
\begin{figure}[htbp]
\centering
\includegraphics[scale=0.65]{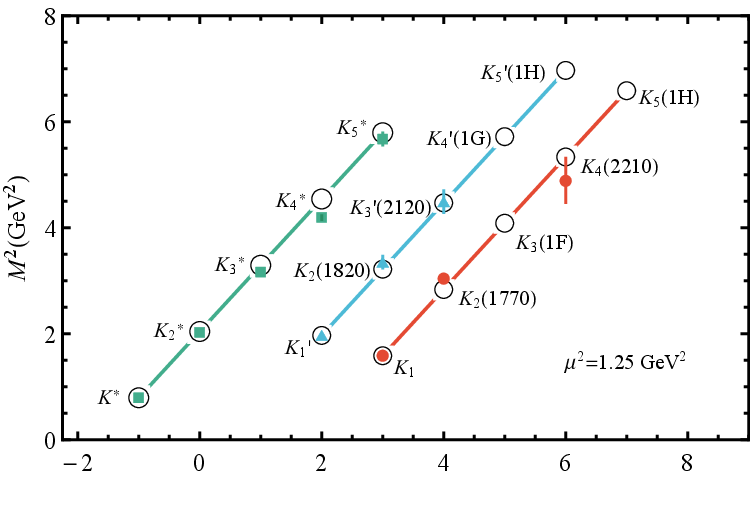}
\caption{The Regge trajectories of the $K_J^*(L=J-1)$, $K_J$, and $K_J^\prime$ families. 
Open circles denote the theoretical values obtained by Eq. \ref{rt}, while filled geometrical shapes represent experimental data.
Vertical lines indicate the uncertainties associated with each kaon's mass.
For the $K^*$ family, the horizontal axis corresponds to $J-2$, whereas for $K_J^\prime$ family, the horizontal axis represents $J+1$, 
and for $K_J$ kaons, the horizontal axis corresponds to $J+2$, where and $J$ denotes the angular momentum of the corresponding meson. 
}
\label{Regge}
\end{figure}

The investigation of the light meson spectrum can be effectively carried out through the analysis of Regge trajectories \cite{Anisovich:2000kxa, Chew:1962eu}.
The masses and total angular momentum of light mesons that belong to the same meson family satisfy the following relation 

\begin{eqnarray}
M^2=M_0^2+(J-1)\mu^2, \label{rt}
\end{eqnarray}
where {$M_0$} denotes the experimental mass of the ground state, 
$J$ denotes the total angular momentum for the corresponding meson with mass $M$, and $\mu^2$ represents the trajectory slope. 
For the $K^*_J(L=J-1)$, $K_J$, and $K^\prime_J$ families, we adopt a slope of  $\mu^2=1.25$ $\text{GeV}^2$, as depicted in Fig.  \ref{Regge}.

\subsection{The modified Godfrey-Isgur model}

The MGI model was proposed in \red{Ref.}  \cite{Song:2015nia}, grounded on the Godfrey-Isgur (GI) model as detailed in \cite{Godfrey:1985xj}.
In the MGI model, the \red{Hamiltonian} of quark and antiquark system is
\begin{equation}\label{hh}
\tilde{H}=\sum_i{({m_i^2+\mathbf{p}_i^2})}^{1/2}+\tilde{V}^{\mathrm{eff}},
\end{equation}
where $m_i$ denotes the mass of the quark (or antiquark), and $\mathbf{p_{i}}$ denotes the three-momentum of quark (or antiquark), respectively. The effective potential of the $q\bar{q}$ interaction, denoted as $\widetilde{V}^{\mathrm{eff}}$, encompasses a short-range $\gamma^{\mu}\otimes\gamma_{\mu}$ one-gluon-exchange interaction and $1\otimes1$ linear confinement interaction, and is expressed as
\begin{equation}
\widetilde{V}^{\mathrm{eff}}=\widetilde{G}_{12}+\widetilde{V}^{\mathrm{cont}}+\widetilde{V}^{\mathrm{tens}}+\widetilde{V}^{\mathrm{so(v)}}+\widetilde{S}_{12}(r)+\widetilde{V}^{\mathrm{so(s)}},
\label{V}
\end{equation}
which includes the Coulomb term ($\widetilde{G}_{12}$), the contact term ($\widetilde{V}^{\mathrm{cont}}$), the tensor ($\widetilde{V}^{\mathrm{tens}}$), the vector spin-orbit term ($\widetilde{V}^{\mathrm{so(v)}}$), the screened confinement term [$\widetilde{S}_{12}(r)$], and the scalar spin-orbit interaction term ($\widetilde{V}^{\mathrm{so(s)}}$).

The spin-independent terms of the \red{nonrelativistic} potential read
\begin{equation}
\begin{aligned}
\tilde{G}(r)=-\sum_{k=1}^3\frac{4\alpha_k}{3r}\left(\frac{2}{\sqrt{\pi}}\int_0^{\gamma_{k}r} e^{-x^2}dx\right),
\end{aligned}
\end{equation} 
and
\begin{equation}
{S}(r)=\frac{b(1-e^{-\mu r})}{\mu}+c,
\end{equation}
{\color{black}where 
$\alpha_1=0.25$, $\alpha_2=0.15$, $\alpha_3=0.2$
and $\gamma_1=1/2$  GeV, $\gamma_2=\sqrt{10}/2$  GeV, $\gamma_3=\sqrt{1000}/2$ GeV, respectively} \cite{Godfrey:1985xj}, and $\mu=0.0779$ GeV is the screening parameter, whose particular value is required to be determined through comparisons between the theory and experiment \cite{Wang:2024lba}. And $b$ is the confining  parameter, while $c$ is the vacuum constant. Here  $b=0.222$ GeV$^2$ and $c=-0.228$ GeV, which are taken from Ref.~\cite{Wang:2024lba}.
\begin{table}[htbp]
\color{black}
\renewcommand{\arraystretch}{1.5}
\caption{Parameters and their values in this work~\cite{Wang:2024lba}. \label{MGI}}
\begin{center}
\begin{tabular}{cccc}
\hline\hline
Parameter &  value &Parameter &  value  \\
 \midrule[1pt]
$m_u$ (GeV)       &0.162    &{$\sigma_0$ (GeV)}   &{1.791}\\
$m_d$ (GeV)       &0.162    &{$s$ }          &{0.711}\\
$m_s$ (GeV)       &0.377    &$\mu$ (GeV)          &0.0779 \\
$b$ (GeV$^2$)     &0.222    &$c$ (GeV)            &$-0.228$\\
$\epsilon_c$      &-0.137   &$\epsilon_{so(v)}$     &0.0550\\
$\epsilon_{so(s)}$  &0.366    &$\epsilon_t$         &0.493\\
\hline\hline
\end{tabular}
\end{center}
\end{table}

The effective potential $\widetilde{V}^{\mathrm{eff}}$ incorporate relativistic effects, particularly in meson systems, which are implemented in two distinct ways.
First, a smearing function for the meson $q\bar{q}$ is introduced, utilizing nonlocal interactions and novel $\mathbf{r}$ dependencies, and it takes the form
\begin{equation}
\rho_{ij} \left(\mathbf{r}-\mathbf{r^\prime}\right)=\frac{\sigma_{ij}^3}{\pi ^{3/2}}e^{-\sigma_{ij}^2\left(\mathbf{r}-\mathbf{r^
\prime}\right)^2},
\end{equation}
with
\begin{align}
   \sigma_{ij}^2=\sigma_0^2\Bigg[\frac{1}{2}+\frac{1}{2}\left(\frac{4m_im_j}{(m_i+m_j)^2}\right)^4\Bigg]+
  s^2\left(\frac{2m_im_j}{m_i+m_j}\right)^2.
\end{align}
The spin-independent Coulomb term $\widetilde{G}_{12}(r)$ is defined as
\begin{equation}
\begin{split}
\widetilde{G}_{ij}(r)=&\int d^3{\bf r}^\prime \rho_{ij}({\bf r}-{\bf r}^\prime)G(r^\prime)
=-\sum\limits_{k=1}^3\frac{4\alpha_k }{3r}{\rm erf}(\tau_{kij}r),
\end{split}
\end{equation}
where the values of $\tau_{kij}$ are obtained by
\begin{equation}
\tau_{kij}=\frac{1}{\sqrt{\frac{1}{\sigma_{ij}^2}+\frac{1}{\gamma_k^2}}}.
\end{equation}
The screened confinement term $\widetilde{S}_{12}(r)$ is expressed as
\begin{eqnarray}
\widetilde{S}_{12}(r)&=& \int d^3 {\bf{r}}^\prime
\rho_{12} ({\bf r}-{\bf r}^\prime)S(r^\prime)\nonumber\\
&=& \frac{b}{\mu r}\Bigg[r+e^{\frac{\mu^2}{4 \sigma^2}+\mu r}\frac{\mu+2r\sigma^2}{2\sigma^2}\Bigg(\frac{1}{\sqrt{\pi}}
\int_0^{\frac{\mu+2r\sigma^2}{2\sigma}}e^{-x^2}dx-\frac{1}{2}\Bigg) \nonumber\\
&&-e^{\frac{\mu^2}{4 \sigma^2}-\mu r}\frac{\mu-2r\sigma^2}{2\sigma^2}\Bigg(\frac{1}{\sqrt{\pi}}
\int_0^{\frac{\mu-2r\sigma^2}{2\sigma}}e^{-x^2}dx-\frac{1}{2}\Bigg)\Bigg]+c.  \nonumber \\
\label{Eq:pot}
\end{eqnarray}
Secondly, owing to relativistic effects, the general potential must depend on the mass center of the interacting quarks.
Momentum-dependent factors, which are unity in the nonrelativistic limit, are applied as follows:
\begin{equation}
\tilde{G}_{12}(r)\to \tilde{G}_{12}=\left(1+\frac{\mathbf{p}^2}{E_1E_2}\right)^{1/2}\tilde{G}_{12}(r)\left(1 +\frac{\mathbf{p}^2}{E_1E_2}\right)^{1/2}.
\end{equation}
where  $E_{1}, E_{2}$ are the energies of the two constituent quarks.
For the spin-dependent terms, the semirelativistic corrections are applied as follows:
\begin{equation}
\label{vsoij}
  \tilde{V}^i_{\alpha \beta}(r)\to\tilde{V}^i_{\alpha \beta}= \left(\frac{m_\alpha m_\beta}{E_\alpha E_\beta}\right)^{1/2+\epsilon_i} \tilde{V}^i_{\alpha \beta}(r)\left(\frac{m_\alpha m_\beta}{E_\alpha E_\beta}\right)^{1/2+\epsilon_i},
\end{equation}
where $\tilde{V}^i_{\alpha \beta}(r)$ delegate the contact, tensor, vector spin-orbit, and scalar spin-orbit terms, and $\epsilon_i=\epsilon_c$, $\epsilon_t$, $\epsilon_{\rm so(v)}$, and $\epsilon_{\rm so(s)}$ impacts the potentials $\widetilde{V}^{\mathrm{cont}}$, $\widetilde{V}^{\mathrm{tens}}$, $\widetilde{V}^{\mathrm{so(v)}}$, and $\widetilde{V}^{\mathrm{so(s)}}$, respectively \cite{Wang:2021abg}. 
The explicit forms of these spin-dependent terms are given by:
\begin{equation}
\begin{split}
\widetilde{V}^{\mathrm{cont}}=\frac{2{\bf S}_1\cdot{\bf S}_2}{3m_1m_2}\nabla^2\widetilde{G}_{12}^c,
\end{split}
\end{equation}

\begin{equation}\label{Vtens}
\begin{split}
\widetilde{V}^{\mathrm{tens}}=&-\left(\frac{3{\bf S}_1\cdot{\bf r}{\bf S}_2\cdot{\bf r}/r^2-{\bf S}_1\cdot{\bf S}_2}{3m_1m_2}\right)\left(\frac{\partial^2}{\partial r^2}-\frac{1}{r}\frac{\partial}{\partial r}\right)\widetilde{G}_{12}^t,
\end{split}
\end{equation}
\begin{equation}\label{Vsov}
\begin{split}
\widetilde{V}^{\mathrm{so(v)}}=&\frac{{\bf S}_1\cdot {\bf L}}{2m_1^2}\frac{1}{r}\frac{\partial\widetilde{G}_{11}^{\rm so(v)}}{\partial r}+\frac{{\bf S}_2\cdot {\bf L}}{2m_2^2}\frac{1}{r}\frac{\partial\widetilde{G}_{22}^{\rm so(v)}}{\partial r}
\\
 &
+\frac{({\bf S}_1+{\bf S}_2)\cdot {\bf L}}{m_1m_2}\frac{1}{r}\frac{\partial\widetilde{G}_{12}^{\rm so(v)}}{\partial r},\\
\end{split}
\end{equation}

\begin{equation}\label{Vsos}
\begin{split}
\widetilde{V}^{\mathrm{so(s)}}=&-\frac{{\bf S}_1\cdot {\bf L}}{2m_1^2}\frac{1}{r}\frac{\partial\widetilde{S}_{11}^{\rm so(s)}}{\partial r}-\frac{{\bf S}_2\cdot {\bf L}}{2m_2^2}\frac{1}{r}\frac{\partial\widetilde{S}_{22}^{\rm so(s)}}{\partial r}.\\
\end{split}
\end{equation}
As an example, the semirelativistic correction to the contact term following Eq.~\eqref{vsoij} reads:
\begin{equation}
\begin{split}
\widetilde{V}^{\mathrm{cont}}=&\left(\frac{m_1 m_2}{E_1 E_2}\right)^{1/2+\epsilon_c}\frac{2{\bf S}_1\cdot{\bf S}_2}{3m_1m_2}\nabla^2\widetilde{G}_{12}(r)\left(\frac{m_1 m_2}{E_1 E_2}\right)^{1/2+\epsilon_c}\\
=&\left(\frac{m_1 m_2}{E_1 E_2}\right)^{1/2+\epsilon_c}\frac{2{\bf S}_1\cdot{\bf S}_2}{3m_1m_2}\nabla^2\left(-\sum\limits_k\frac{4\alpha_k }{3r}{\rm erf}(\tau_{k12}r)\right)\\
&\left(\frac{m_1 m_2}{E_1 E_2}\right)^{1/2+\epsilon_c}. \nonumber
\end{split}
\end{equation}
The values of parameters $m_u$, $m_d$, $m_s$, $b$, $\epsilon_c$, $\epsilon_{so(s)}$, $\sigma_0$, $s$, $\mu$, $c$, $\epsilon_{so(v)}$ and $\epsilon_t$ are taken from \red{Ref.} \cite{Wang:2024lba} as listed in Table \ref{1}.

By diagonalizing the Hamiltonian matrix depicted in Eq. (\ref{hh}) within the framework of the simple harmonic oscillator (SHO) basis, the mass spectrum and corresponding spatial wave functions of the meson can be obtained. These wave functions can  be applied to the study of strong decay processes. 

 \subsection{A brief review of the \text{$^3P_0$} model}\label{3p0}

The $^3P_0$ model was originally proposed by Micu \cite{Micu:1968mk} and subsequently developed by the Orsay group \cite{LeYaouanc:1972ae, LeYaouanc:1973xz, LeYaouanc:1974mr, LeYaouanc:1977gm, LeYaouanc:1977ux}. This model is broadly employed in the analysis of OZI-allowed two-body strong decays of mesons \cite{vanBeveren:1982qb, Titov:1995si, Ackleh:1996yt, Blundell:1996as, Bonnaz:2001aj, Zhou:2004mw, Lu:2006ry, Zhang:2006yj, Luo:2009wu, Sun:2009tg, Liu:2009fe, Sun:2010pg, Rijken:2010zza, Ye:2012gu, Wang:2012wa, He:2013ttg, Sun:2013qca,  Wang:2014sea,Pang:2014laa, Pang:2015eha,  Chen:2015iqa, Pang:2017dlw,Pang:2018gcn,Pang:2019ttv,Wang:2020due,  Wang:2021abg, feng:2021igh, Feng:2022hwq,  Wang:2022juf, Wang:2022xxi, Li:2022khh, Li:2022bre, Wang:2024yvo}.
The transition operator $\mathcal{T}$ describes a quark-antiquark pair (denoted by indices $3$ and $4$) creation from vacuum with $J^{PC}=0^{++}$.
For the process $A\to B+C$, 

\begin{eqnarray}
\langle BC|\mathcal{T}|A \rangle = \delta ^3(\mathbf{P}_B+\mathbf{P}_C)\mathcal{M}^{{M}_{J_{A}}M_{J_{B}}M_{J_{C}}}.
\end{eqnarray}
$\mathcal{T}$ can be written as \cite{Wang:2019jch}
{\begin{align}\label{gamma}
\mathcal{T} = & -3\gamma \sum_{m}\langle 1m;1~-m|00\rangle\int d \mathbf{p}_3d\mathbf{p}_4\delta ^3 (\mathbf{p}_3+\mathbf{p}_4) \nonumber \\
 & ~\times \mathcal{Y}_{1m}\left(\frac{\textbf{p}_3-\mathbf{p}_4}{2}\right)\chi _{1,-m}^{34}\phi _{0}^{34}
\left(\omega_{0}^{34}\right)_{ij}b_{3i}^{\dag}(\mathbf{p}_3)d_{4j}^{\dag}(\mathbf{p}_4),
\end{align}}
where $\mathcal{Y}_l^m(\bf{p})\equiv$ $p^lY_l^m(\theta_p,\phi_p)$ represents the solid harmonics. 
$\chi$, $\phi$, and $\omega$ denote the spin, flavor, and color wave functions respectively. 
Subindices $i$ and $j$ denote the color index associated with the $q\bar{q}$ pair. The amplitude $\mathcal{M}^{{M}_{J_{A}}M_{J_{B}}M_{J_{C}}}$ of the decay process is defined with the transition operator $\mathcal{T}$. 

The two decay amplitudes can be related by the Jacob-Wick formula \cite{Jacob:1959at}
\begin{equation}
\begin{aligned}
\mathcal{M}^{JL}(\mathbf{P}) = &\frac{\sqrt{4\pi(2L+1)}}{2J_A+1}\sum_{M_{J_B}M_{J_C}}\langle L0;JM_{J_A}|J_AM_{J_A}\rangle \\
    &\times \langle J_BM_{J_B};J_CM_{J_C}|{J_A}M_{J_A}\rangle \mathcal{M}^{M_{J_{A}}M_{J_B}M_{J_C}}.
\end{aligned}	
\end{equation}
Finally, the general form of the decay width can be expressed as
\begin{eqnarray}
\Gamma&=&\frac{\pi}{4} \frac{|\mathbf{P}|}{m_A^2}\sum_{J,L}|\mathcal{M}^{JL}(\mathbf{P})|^2,
\end{eqnarray}
where $m_{A}$ denotes the mass of the initial state $A$, and
$\bf{P}$ is the three-momentum of initial meson B in the rest frame of the meson A. 

The dimensionless parameter $\gamma$ characterizes the strength of $q\bar{q}$ pair creation from the vacuum in this model.  
In this work, we take the $\gamma$ value to be $6.31$ for high spin kaons, which fit the widths of $F$-wave $K_4^*(2405)$ and $G$-wave $K_5^*(2380)$ states.
Furthermore, for the spatial wave functions of the kaons under consideration, we utilize the MGI model, as detailed in the preceding section.

There are two types of mixing schemes in the light mesons. One is the spin mixing. For nature states of kaons, they should be the superpositions of the pure states $n^{1}L_J$ and $n^{3}L_J$, where quantum number $L=J$. 
The spin mixing scheme for the kaons can be expressed as
\begin{equation}\label{anglek1}
\left( \begin{matrix}
	|K(nL)\rangle \\
	|K^\prime(nL)\rangle \\
\end{matrix}\right) =
\left( \begin{matrix}
	\textrm{$\cos\theta_{nL}$} & \textrm{$\sin\theta_{nL}$} \\
	\textrm{$-\sin\theta_{nL}$} & \textrm{$\cos\theta_{nL}$} \\
\end{matrix}\right)
\left( \begin{matrix}
	|K(n^1L_L)\rangle \\
	|K(n^3L_L)\rangle \\
\end{matrix}\right),
\end{equation}
where $\theta_{nL}$ is known as the mixing angle between the $K(n^1L_L)$ and the $K(n^3L_L)$ states. 
The value of the mixing angle $\theta_{1P}=-34^\circ$ \cite{Cheng:2013cwa}. For other mixing angles  
$\theta_{nP}=-35.3^\circ(n>1)$, 
$\theta_{nD}=-39.2^\circ$, 
$\theta_{nF}=-40.9^\circ$, 
$\theta_{nG}=-41.8^\circ$, and 
$\theta_{nH}=-42.4^\circ$ \cite{bokade2025bc}.

The other mixing scheme is flavor mixing.
For flavor wave functions of the isoscalar  light mesons may have the mixing form
\begin{equation}\label{mixingns}
\left( \begin{matrix}
	X \\
	X^\prime \\
\end{matrix}\right) =
\left( \begin{matrix}
	\textrm{$\cos\phi_{x}$} & \textrm{$\sin\phi_{x}$} \\
	\textrm{$-\sin\phi_{x}$} & \textrm{$\cos\phi_{x}$} \\
\end{matrix}\right)
\left( \begin{matrix}
	|n\bar{n}\rangle \\
	|s\bar{s}\rangle \\
\end{matrix}\right),
\end{equation}
where $X$ and $X^\prime$ are the two isoscalar mesons (such as $\eta$ and $\eta^\prime$), 
$\phi_x$ is the mixing angle in the quark flavor scheme, and $n\bar{n}=(u\bar{u}+d\bar{d})/\sqrt{2}$. 
In this work, we adopted $\phi_{f_1(1P)}=-24^\circ$ and $\phi_{h_1(1P)}=4.4^\circ$. 
For $\eta$ meson, we take $\phi_{\eta(1S)}=-39.3^\circ$.   
The $\phi_x$ for other isoscalar light mesons are considered as pure $n\bar{n}$ or $s\bar{s}$ states.

The mesons  $\omega$, $\rho$, $\eta^\prime$, $\phi$, $b_1$, $h_1$, $h_1^\prime$,  $a_0$, $a_1$, $a_2$, $f_1$,  $K^*$, $K_1$, ${K_1}^\prime$,  and $K_2^*$ correspond to the mesons  $\omega(782)$, $\rho(770)$, $\eta^\prime(958)$, $\phi(1020)$, $b_1(1235)$, $h_1(1170)$, $h_1(1380)$, $a_0(1450)$, $a_1(1260)$, $a_2(1320)$, $f_1(1285)$,   $K^*(892)$, $K_1(1270)$, ${K_1}(1400)$, and  $K_2^*(1430)$, respectively.
Additional decay modes, including $K^*(3S)$, $K_3^*(2D)$, $f_2(2P)$, $K_2^*(3P)$, $\rho_2(1D)$, and others, correspond to theoretical states that have not yet been observed experimentally. 
These states are listed in Tables~\ref{K31F}, \ref{K32F}, \ref{K3star}, \ref{K41G}, \ref{K42G}, \ref{K4star}, \ref{K51H}, and \ref{K5star}, and their masses are predicted using the MGI model. Decays with partial widths smaller than 1 MeV are omitted from the tables for clarity.

\section{Numerical results}\label{3}

\begin{table*}[htbp]
\renewcommand{\arraystretch}{1.3}
\centering
\caption{The \red{mass} spectrum of the high spin kaons. The results of “this work" \red{do not} take into account spin-mixing. The unit of mass is MeV. “Exp." represents the experimental value. \label{mass}}
\vspace{-0.2cm}
\[\begin{array}{cccccccccccc}
\hline
\hline
\text{States}&J^{P}&n^{2S + 1}L_J&\text{this work}&\text{GI}$~\cite{Steph:1985ff}$&\text{Segovia}$~\cite{Taboada-Nieto:2022igy}$&\text{Ebert}$~\cite{Ebert:2009ub}$& \text{Exp.} \\\midrule[1pt]
K_3(1F)         &3^{+}& 1^1F_3  &2066  &2130   &2047   &2009  & \cdots\\
K_3^\prime(2120)&3^{+}& 1^3F_3  &2064  &2146   &2132   &2080   &2119\pm13^{+45}_{-12}~$\cite{COMPASS:2025wkw}$\\
K_3^*(1780)       &3^{-}& 1^3D_3  &1794  &1780   &1810   &1789   &1779\pm8$~\cite{ParticleDataGroup:2024cfk}$ \\ 
K_3(2320)         &3^{+}& 2^1F_3  &2347  &2524   &2340   &2384   &2324\pm24$~\cite{ParticleDataGroup:2024cfk}$  \\
K_3^\prime(2F)    &3^{+}& 2^3F_3  &2347  &2539   &2387   &\cdots &\cdots \\ \hline
K_4(2210)         &4^{-}& 1^1G_4  &2307  &2422   &2270   &2255   &2210\pm40^{+80}_{-30}~$\cite{COMPASS:2025wkw}$ \\
K_4^\prime(1G)    &4^{-}& 1^3G_4  &2306  &2435   &2337   &2285   &\cdots   \\
K_4^*(2045)       &4^{+}& 1^3F_4  &2075  &2097   &2080   &2096   &2060\pm5^{+11}_{-3}~$\cite{COMPASS:2025wkw}$, 2048^{+8}_{-9}$~\cite{ParticleDataGroup:2024cfk}$  \\  
K_4(2500)         &4^{-}& 2^1G_4  &2538  &2778   &2476   &2575  &2490\pm20$~\cite{ParticleDataGroup:2024cfk}$      \\
K_4^\prime(2G)    &4^{-}& 2^3G_4  &2538  &2790   &2510   &\cdots             &\cdots \\ \hline
K_5(1H)           &5^{+}& 1^1H_5  &2509  &2682   &2442   &\cdots          &\cdots \\
K_5^\prime(1H)    &5^{+}& 1^3H_5  &2509  &2693   &2489   &\cdots          &\cdots  \\
K_5^*(2380)       &5^{-}& 1^3G_5  &2309  &2379   &2291   &\cdots &2382\pm24$~\cite{ParticleDataGroup:2024cfk}$   \\  
 \hline
 \hline
\end{array}\]
\end{table*}

Regge trajectory analysis, as depicted in Fig. \ref{Regge}, reveals that $K_1$, $K_2(1770)$, $K_3(1F)$, and $K_4(2210)$ lie on linear trajectories in the ($J, M^2$)-planes, as do $K_1^\prime$, $K_2(1820)$, $K_3^\prime(2120)$ [was called $K_3(2120)$ in the COMPASS observation \cite{COMPASS:2025wkw}], $K_4^\prime(1G)$, and $K_5^\prime(1H)$.
Thereafter, we utilized the MGI model to determine the mass spectrum of $J^{PC}=3^\pm, 4^\pm$, and $5^\pm$ kaons, which is detailed in Table \ref{mass}.
Furthermore, employing the spatial wave functions of mesons obtained from the MGI model as input, we calculated the OZI-allowed two-body strong decays with initial masses taking both experimental and theoretical values of $J^{PC}=3^\pm, 4^\pm$, and $5^\pm$ kaons, results of which are provided in Tables \ref{K3star}, \ref{K31F}, \ref{K32F}, \ref{K41G}, \ref{K42G}, \ref{K4star}, \ref{K51H}, and \ref{K5star}.
In addition, we investigated the correlation between the spin-mixing properties of strange mesons and the mixing angles of $K_3^\prime(2120)$, $K_3(2320)$, and $K_4(2210)$, as depicted in Figs. \ref{K32120}, \ref{K32320}, and \ref{K42210}. 
For the subsequent discussions, the experimental value is adopted as the initial-state mass unless stated otherwise.

\subsection{High spin kaons with $J^{P}=3^{\pm}$}

For $J^{P}=3^{-}$, only $K_3^*(1780)$ is listed in the PDG. The first recorded observation of $K_3^*(1780)$ dates back to 1976 \cite{Brandenburg:1975ft}, where its spin parity was established. In the same year, it was rediscovered in the reaction $K^+ p \to K^0_s \pi^+ p$ at an energy of 10 GeV, with a mass of $1779 \pm 11$ MeV and a width of $135 \pm 22$ MeV, respectively \cite{Baldi:1976ua}. 
In 1978, Chung and collaborators reported the discovery in $K^-\pi^+$ mass sprectrum with a mass of $1786\pm8$ MeV and a width of $95\pm31$ MeV resulting from the reaction $K^-p \to K^-\pi^+n$ at an energy of $6$ GeV. This observation was made using the Brookhaven National Laboratory's multiparticle spectrometer \cite{Chung:1977ji}. 
Subsequently, research on $K_3^*$ has accumulated, with an expanding body of studies and measurements having been carried out  \cite{Konigs:1978at, Etkin:1980me, Toaff:1981yk, Aston:1980bp, Cleland:1982td, BIRMINGHAM-CERN-GLASGOW-MICHIGANSTATE-PARIS:1982skm, Birmingham-CERN-Glasgow-MichiganState-Paris:1984ppi, Aston:1986jb, Aston:1987ir, Aston:1987ey}. 
The latest measurement of $K_3^*$ was performed in 2020 through a partial wave analysis of $\psi(3686)\to K^{+}K^{-}\eta$ by the BESIII Collaboration \cite{BESIII:2019dme}. 
The mass and width of the $K_3^*(1780)$ resonance were determined to be $1813 ^{{+15}{+65}}_{{-15}{-16}}$ MeV and $191^{{+43}{+3}}_{{-37}{-81}}$ MeV, respectively  \cite{BESIII:2019dme}. 
The $K_3^*(1780)$ is an established $1^3D_3$ state. 
The partial decay behaviors for $K_3^*(1780)$ as the $1^3D_3$ state are summarized in Table \ref{K3star}. 
The dominant decay channels of $K_3^*(1780)$ \purple{is $\rho K^*$.}
$\omega K^*$, $\pi K$, $\pi K^*$, $K \rho$, $K\eta$, $K\omega$, $\pi K_2^*$, and $\pi K_1$ are important decay channels. 
The ratio $\Gamma_{\pi K}/\Gamma_{\pi K^*}$ is about 1, which is in consistent with the experimental value $0.94\pm0.2$ \cite{ParticleDataGroup:2024cfk}. Besides, the branching ratio of the $\pi K_2^*$ is 0.014, which is in  consistent with the experimental value less than 0.16 \cite{ParticleDataGroup:2024cfk}.  
In Table \ref{K3star}, the decay results obtained using the theoretical value (1794 MeV) of $K_3^*(1780)$ are also listed. Since the experimental and theoretical masses of $K_3^*(1780)$ differ only slightly, the corresponding differences in branching ratios and partial widths are small.

\begin{table}[htbp]
\renewcommand{\arraystretch}{1.3}
\centering
\color{black}
\caption{The decay information of $K_3^*(1780)(1^3D_3)$. 
The unit of the width is MeV. The abbreviation “Br” denotes the branching ratio. The data in $\{~\}$ means the width calculated using the theoretical mass of the initial state.  \label{K3star} }
\[\begin{array}{cccc}
\hline
 \hline
 \text{Channel} & \text{Br(\%)}   & \text{Exp.}~$\cite{ParticleDataGroup:2024cfk}$   & \text{Value} \\
 \midrule[1pt]
 \rho K^*     & 43.2\{44.3\}    & \cdots         & 48.7\{57.8\} \\
 \omega K^*   & 13.4\{14.0\}    & \cdots         & 15.1\{18.2\} \\
 \pi K        & 12.8\{11.7\}    &0.188\pm0.01   & 14.4\{15.3\} \\
 \pi K^*      & 10.5\{9.99\}    &0.2\pm0.05     & 11.8\{13.0\} \\
 K\rho        & 8.51\{8.27\}    &0.31\pm0.09    & 9.59\{10.8\} \\
 K\eta        & 4.26\{4.02\}    &0.3\pm0.13     & 4.8\{5.25\} \\
 K\omega      & 2.73\{2.66\}    & \cdots         & 3.08\{3.47\} \\
 \pi K_2^*    & 1.35\{1.53\}    & <0.16         & 1.52\{2\} \\
 \pi K_1      & 1.29\{1.32\}    &  \cdots        & 1.45\{1.72\} \\
 \multicolumn{4}{c}{\cdots} \\ \hline
 \text{Total} & \text{100$\{$100$\}$} & 161\pm17 & \text{113$\{$130$\}$} \\
\hline
 \hline
\end{array}\]
\end{table}

Kaons with $J^P = 3^+$ quantum number are characterized by either a spin singlet state or a spin triplet state. Nonetheless, the observed natural states are superpositions of both the spin singlet and spin triplet states. We adopted the mixing scheme that was introduced in \red{Sec.} \ref{3p0}. 
In the following, we will present a comprehensive overview of $J^P=3^+$ strange mesons.

The $J^P=3^+$ state $K_3(1F)$ has not yet been observed experimentally. 
Our prediction for the mass of $K_3(1F)$ is 2066 MeV, which is lower than the value derived from the GI model  
\cite{Steph:1985ff} and slightly higher than the value reported in \red{Refs.} \cite{Taboada-Nieto:2022igy} and \cite{Ebert:2009ub}. 
For $K_3(1F)$ and $K_3^\prime(2120)$, we use the same experimental value as input for the calculation.
The total decay width of $K_3(1F)$ is calculated to be 248 MeV. Meanwhile, we calculated the width of the $K_3(1F)$ state using theoretical mass as shown in Table \ref{K31F}. The result indicates that since the theoretical mass of $K_3(1F)$ is smaller than the experimental mass, the phase spaces of the $\rho K_1$ and $\omega K_1$ decay channels are compressed. Consequently, the width derived from the theoretical mass is smaller than that obtained using the experimental mass.
The most important decay channels of $Ka_2$, $\rho K^*$, $\rho K_1$, $K f_2$, and $\pi K^*$ can be employed to identify the $K_3(1F)$ in experiment.
Given that $a_2$ decays predominantly to $\rho\pi$,  
$K^*$ mainly decays into $K\pi$, and 
$\rho$ almost exclusively decays into $\pi\pi$,
$K3\pi$ emerges as the dominant final state of  $K_3(1F)$.
Furthermore, several other decay modes, including $K \rho$, $\pi K_1$, $\omega K^*$, $\omega K_1$, and $\pi K^*(1410)$, contribute significantly to the overall width, akin to the aforementioned channels. For comprehensive details, refer to Table \ref{K31F}.

\begin{figure}[htbp]
\centering
\includegraphics[scale=1]{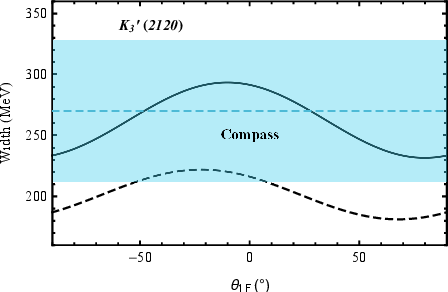}
\caption{{\color{black}The $\theta$ dependence of the total decay width for $K_3^\prime(2120)$, with the corresponding experimental data \cite{COMPASS:2025wkw} (represented by a blue band) being presented alongside our theoretical calculation for comparison. The solid line corresponds to the result obtained using the experimental mass, while the dashed line denotes that obtained with the theoretical mass derived from the MGI model.}}

\label{K32120}
\end{figure}

The dependence of the total width of $K_3^\prime(2120)$ on the mixing angle $\theta_{1F}$ is illustrated in Fig.~\ref{K32120}.
The calculated total width using the experimental mass of $K_3^\prime(2120)$ overlaps with the experimental value  over the entire mixing angle range \cite{COMPASS:2025wkw}.  When the theoretical mass is used for the initial state,  the total width of $K_3^\prime(2120)$ is lower than that obtained with the experimental mass, owing to phase space compression of $K^*h_1$, $Kf_2^\prime(1525)$, $\rho K_1$, and $\omega K_1$.
When the mixing angle takes values of $\theta_{1F}=-48^\circ$ or $\theta_{1F}=27^\circ$, the  calculated total width of $K_3^\prime(2120)$ is in good agreement with the experimental value of 270 MeV \cite{COMPASS:2025wkw}.   
Additionally, Table \ref{K31F} provides a comprehensive list of partial decay modes of $K_3^\prime(2120)$, which is assigned as the $1F$ state at a specific mixing angle of $\theta_{1F}=-40.9^ \circ$.
The primary decay channels encompass $\pi K_3^*(1780)$, $\pi K_2^*$, $\rho K^*$, $K\rho$, and $K^* h_1$ with respective widths of 71 MeV, 48 MeV, 27 MeV, 21 MeV, and 12 MeV. 
The total decay width of $K_3^\prime(2120)$ we calculated is 277 MeV, which is consistent with the value $270\pm30^{+40}_{-30}$ MeV measured by the \purple{COMPASS Collaboration} \cite{COMPASS:2025wkw}. Besides, the $K_3^\prime(2120)$ was observed in the $\pi K_3^*(1780)$ and $\pi K_2^*$ decays, which is consistent with our calculation result, where the branching ratios for $\pi K_3^*(1780)$ and $\pi K_2^* $ are about 0.26 and 0.17, respectively.

\begin{table*}[htbp]
\renewcommand{\arraystretch}{1.5}
\centering
\color{black}
\caption{The decay information of $K_3(1F)$ and $K_3^\prime(2120)(1F)$ with the mixing angle $\theta_{1F} = -40.9^\circ$. The unit of the width is MeV. The data in $\{~\}$ \red{mean} the width calculated using the theoretical mass of the initial state. \label{K31F}}
\[\begin{array}{ccc|ccccc}
\hline
 \hline
  \multicolumn{3}{c|}{K_3(1F)}&\multicolumn{3}{c}{K_3^\prime(2120), \Gamma_{\text{Exp}}=270\pm30^{+40}_{-30}~$\cite{COMPASS:2025wkw}$}
  \\
\hline
 \text{Channel} & \text{Br(\%)} & \text{Value} & \text{Channel} & \text{Br(\%)} & \text{Value} \\
 \midrule[1pt]
  Ka_2          & 28.1\{34.9\}     & 69.6\{64.7\}      & \pi K_3^*(1780)   & 25.4\{30.6\}    & 70.6\{66.7\} \\
 \rho K^*       & 11\{11.6\}      & 27.3\{21.4\}      & \pi K_2^*         & 17.1\{22.7\}    & 47.5\{49.4\} \\
 \rho K_1       & 10.2\{2.08\}    & 25.3\{3.85\}      & \rho K^*            & 9.64\{9.56\}   & 26.7\{20.8\} \\
 Kf_2           &9.34\{12.3\}     & 23.1\{22.7\}      & K \rho              & 7.44\{7.31\}   & 20.6\{15.9\} \\
 \pi K^*        & 8.92\{9.44\}   & 22.1\{17.5\}      & K^*h_1              & 4.29\{0.223\}   & 11.9\{0.485\} \\
 K \rho         & 5.44\{7.62\}   & 13.5\{14.1\}      & Kf^\prime_2(1525)   &4.22\{1.75\}  & 11.7\{3.81\} \\
 \pi K_1        & 4.09\{4.13\}   & 10.1\{7.66\}      & \pi K^*             & 4.18\{5.74\}   & 11.6\{12.5\} \\
 \omega K^*     & 3.6\{3.76\}    & 8.91\{6.97\}      & \rho K_1            &3.64\{0.693\}   & 10.1\{1.51\} \\
 \omega K_1     & 3.05\{0.295\}  & 7.54\{0.547\}     & Ka_1                & 3.32\{2.63\}   & 9.21\{5.72\} \\
\pi K^*(1410)   & 2.85\{2.26\}   & 7.05\{4.19\}      & \omega K^*          &3.14\{3.11\}   & 8.72\{6.77\}\\
 \eta K_2^*     & 2.65\{1.83\}   & 6.57\{3.39\}      & \eta K_2^*          & 2.73\{2.22\}   & 7.58\{4.83\} \\
 K \omega       & 1.83\{2.55\}   & 4.53\{4.73\}      & K \omega            & 2.43\{2.38\}   & 6.74\{5.18\} \\
 \pi K_2^*      & 1.81\{1.5\}    & 4.47\{2.77\}      &K \phi               & 1.68\{1.81\}   & 4.65\{3.93\} \\
 \eta K^*       & 1.55\{1.65\}   & 3.85\{3.05\}      & \pi K_1             & 1.61\{1.60\}  & 4.47\{3.48\} \\
\pi K_0^*(1430) & 0.783\{0.722\} & 1.94\{1.34\}      & \eta K^*            & 1.48\{1.80\}   & 4.12\{3.91\} \\
 \pi K_1^\prime & 0.756\{0.718\} & 1.87\{1.33\}      & \pi K^*(1410)       & 1.32\{1.99\}   & 3.67\{4.34\} \\
 Kb_1           & 0.547\{0.44\}  & 1.35\{0.493\}     & Ka_2                & 1.22\{1.07\}   & 3.39\{2.32\} \\
\pi K_2(1770)   & 0.546\{0.266\} & 1.35\{0.816\}     & \omega K_1          & 1.08\{0.0984\} & 3.00\{0.214\} \\
 K^*\phi        & 0.483\{0.32\}  & 1.20\{0.593\}     & Kb_1                & 0.651\{0.500\}   & 1.81\{1.09\} \\
\multicolumn{3}{c|}{\cdots}                          & Kf_2                & 0.527\{0.468\}   & 1.46\{1.02\} \\
          &         &                                & Kf_1                & 0.501\{0.359\}   & 1.39\{0.783\} \\
          &         &                                & K^*\phi             & 0.409\{0.258\}   & 1.13\{0.561\}\\
          &         &                                & K^* \eta^\prime     & 0.369\{0.325\}   & 1.02\{0.707\} \\  
          &         &                                &\multicolumn{3}{c}{\cdots}\\
\hline
 \text{Total} & 100\{100\} & 248\{185\} & \text{Total} & 100\{100\} & 277\{218\}\\
 \hline
 \hline
\end{array}\]
\end{table*}

 The $K_3(2320)$ state was first observed in the $p\bar{\Lambda}$ system in 1981, with a mass $M=2320 \pm 30$ MeV and a width $\Gamma \sim 150$  MeV  \cite{Cleland:1980ya}. 
 In 1983, a partial wave analysis interpreted this \purple{structure} as a resonances with spin parity $3^-$, mass $M = 2330 \pm 40$ MeV,  and width  $\Gamma = 150 \pm 30$  MeV, respectively  
\cite{Bari-Birmingham-CERN-Milan-Paris-Pavia:1983nwf}.

\begin{figure}[htbp]
\centering%
\includegraphics[scale=1]{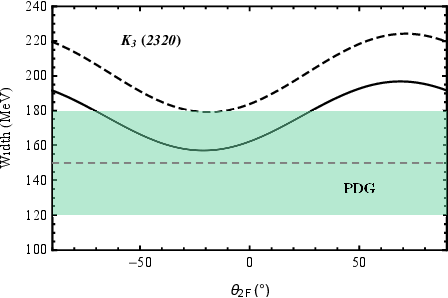}
\caption{{\color{black}The $\theta$ dependence of the total decay width of the $K_3(2320)$, 
 along with the corresponding experimental data~\cite{ParticleDataGroup:2024cfk} (represented by a light green band) which is provided for comparison with our theoretical calculation. The solid line corresponds to the result obtained using the experimental mass, while the dashed line denotes that obtained with the theoretical mass derived from the MGI model.}}
\label{K32320}
\end{figure}

The first excited states of the $K_3(1F)$ and $K_3^\prime(2120)$ are the $K_3(2320)$ and $K_3^\prime(2F)$, respectively. For $K_3(2320)$ and $K_3^\prime(2F)$, the same experimental mass is adopted for the initial states and their OZI-allowed two-body strong decays are summarized in Table \ref{K32F}.
The $\theta_{2F}$ dependence of the total decay width of the $K_3(2320)$ state is depicted in Fig. \ref{K32320}. The theoretical total width of the $K_3(2320)$ is consistent with the experimental value for mixing angles in the interval $[-70^\circ,28^\circ]$ \cite{ParticleDataGroup:2024cfk}. The theoretical mass of $K_3(2320)$ is slightly higher than its experimental counterpart. When the mixing angle is approximately $-20^\circ$, the width derived using the theoretical mass of $K_3(2320)$ as input can reach the upper limit of the PDG experimental value \cite{ParticleDataGroup:2024cfk}.

The determined total width of $K_3(2320)$ is 161 MeV, which corresponds to an angle $\theta_{2F}$ of $-40.9^ \circ$. 
This value lies within the experimental error bounds of $150\pm30$ MeV \cite{ParticleDataGroup:2024cfk}. The most important decay modes for $K_3(2320)$ include $K a_2(1700)$, $Ka_2$, $\pi K^*(1410)$, $K\rho$, $Kf_2(2P)$, and $\rho K^*$, all of which are instrumental for the precise determination of the $K_3(2320)$ structure.

As a counterpart to the $K_3(2320)$, we have predicted the mass and OZI-allowed \purple{two-body} strong decay behaviors of the $K_3^\prime(2F)$, and presented the results in Tables \ref{mass} and \ref{K32F}. 
Our calculations reveal that the mass and width of $K_3^\prime(2F)$ are 2347 MeV and 192 MeV, respectively. The predicted mass is lower than the values obtained by the GI model \cite{Steph:1985ff} and \red{Ref.} \cite{Taboada-Nieto:2022igy}. The $\pi K_3^*(2D)$, $\pi K_2^*$, $\pi K_3^*(1780)$, $\pi K_2^*(1980)$, and $\pi K_4^*(2045)$ channels are more suitable to investigate the $K_3^\prime(2F)$. Additionally, the $\pi K^*$, $\rho K^*(1410)$, $K \rho(1450)$, $\rho K^*$, $K^* \pi(1300)$ and other modes  contribute significantly to the total decay width.

\begin{table*}[htbp]
\renewcommand{\arraystretch}{1.5}
\centering
\color{black}
\caption{The decay information of $K_3(2320)(2F)$ and $K_3^\prime(2F)$ with the mixing angle $\theta_{2F} = -40.9^\circ$. 
The unit of the width is MeV. \label{K32F}}
\[\begin{array}{ccc|ccccc}
\hline
 \hline
\multicolumn{3}{c|}{K_3(2320), \Gamma_{\text{Exp}}=150\pm30$~\cite{ParticleDataGroup:2024cfk}$}&\multicolumn{3}{c}{K_3^\prime(2F)}
  \\
\hline
 \text{Channel} & \text{Br(\%)} &  \text{Value} &  \text{Channel} &  \text{Br(\%)} & \text{ Value} \\
\midrule[1pt]
 Ka_2(1700) & 15.6\{15.6\}     &25.2\{29\}   & K_3^*(2D)\pi  & 15.2\{15.1\} & 29.2\{32.8\}\\
 Ka_2       & 10.9\{10.6\} & 17.7\{19.7\} & K_2^*\pi  & 10.6\{9.62\} & 20.5\{21\} \\
 K^*(1410)\pi  & 5.74\{5.01\} & 9.27\{9.3\} & K_3^*(1780)\pi  & 10\{9.04\} & 19.3\{19.7\}\\
 K\rho  & 5.07\{4.52\} & 8.19\{8.39\} & K_2^*(1980)\pi  & 9.39\{8.43\} & 18.1\{18.4\}\\
 Kf_2(2P) & 4.89\{5.53\} & 7.9\{10.3\} & K_4^*(2045)\pi  & 7.05\{5.87\} & 13.6\{12.8\}\\
 K^*(1410)\rho  & 4.39\{4.74\} & 7.1\{8.79\} & K^*\pi  & 5.09\{4.54\} & 9.79\{9.89\}\\
 Kf_2 & 4.28\{4.04\} & 6.92\{7.49\} & K^*(1410)\rho & 3.56\{3.93\} & 6.86\{8.56\} \\
 K_1\rho  & 3.52\{3.89\} & 5.68\{7.22\} & K\rho (1450) & 2.81\{3.03\} & 5.4\{6.59\}\\
 K\rho (1450) & 3.33\{2.79\} & 5.38\{5.17\} & K^*\rho  & 2.67\{2.51\} & 5.14\{5.46\}\\
 K^*\rho  & 3.31\{3.08\} & 5.35\{5.71\} & K^*\pi (1300) & 2.58\{2.7\} & 4.97\{5.88\}\\
 K\rho _3(1690) & 2.83\{3.79\} & 4.58\{7.03\} & K_1\rho  & 2.4\{2.55\} & 4.62\{5.55\}\\
 K_2^*\pi  & 2.79\{2.71\} & 4.51\{5.03\} & a_2K^* & 1.93\{1.86\} & 3.71\{4.05\} \\
 K_1^\prime(2P)\pi  & 2.58\{2.49\} & 4.18\{4.62\} & Ka_1(1640) & 1.64\{2.03\} & 3.16\{4.41\}\\
 K(1460)\rho  & 2.13\{2.53\} & 3.45\{4.7\} & Ka_2 & 1.45\{1.48\} & 2.8\{3.23\} \\
 b_1K^* & 2.11\{2.27\} & 3.4\{4.21\} & K_2^*\eta  & 1.29\{1.32\} & 2.48\{2.88\} \\
 a_1K^* & 1.76\{2.08\} & 2.84\{3.86\} & K\omega (1420) & 1.27\{1.33\} & 2.44\{2.89\}\\
 K_2^*\rho  & 1.72\{1.69\} & 2.78\{3.14\} & K^*(1410)\omega  & 1.11\{1.24\} & 2.13\{2.7\}\\
 K\omega  & 1.68\{1.5\} & 2.72\{2.79\} & K\rho _2(1D) & 1.07\{1.1\} & 2.05\{2.4\}\\
 K\omega _3(1670) & 1.4\{1.82\} & 2.27\{3.37\} & K^*\eta  & 1.03\{0.956\} & 1.99\{2.08\}\\
 K^*(1410)\omega  & 1.37\{1.5\} & 2.21\{2.78\} & K^*(3S)\pi  & 1.03\{1.1\} & 1.98\{2.39\}\\
 K\phi  & 1.23\{0.0236\} & 1.98\{0.0438\} & Kf_0(1500) & 1.02\{1.03\} & 1.96\{2.24\}\\
 K^*(1410)\eta & 1.16\{1.13\} & 1.88\{2.1\} & h_1K^* & 0.967\{1.22\} & 1.86\{2.66\}\\
 K_1\omega  & 1.09\{1.23\}& 1.77\{2.28\} & K^*\eta (1295)& 0.896\{0.923\} & 1.72\{2.01\}\\
 K^*\omega  & 1.09\{1.01\} & 1.76\{1.88\} & K^*\omega  & 0.879\{0.826\} & 1.69\{1.8\} \\
 h_1K^* & 1.07\{1.05\} & 1.72\{1.94\} & b_1K^* & 0.831\{1.29\} & 1.6\{2.81\} \\
 K\omega (1420) & 1.03\{0.839\} & 1.66\{1.56\} & f_2K^* & 0.782\{0.845\} & 1.5\{1.84\}\\
 a_2K^* & 0.838\{0.829\} & 1.35\{1.54\} & a_1K^* & 0.775\{0.855\} & 1.49\{1.86\} \\
 K^*\pi  & 0.746\{0.852\} & 1.21\{1.58\} & K_1\omega  & 0.761\{0.814\} & 1.46\{1.77\} \\
 K^*(1680)\pi  & 0.718\{0.608\} & 1.16\{1.13\} & K_2^*\rho  & 0.712\{0.844\} & 1.37\{1.84\}\\
 K_2(1770)\pi  & 0.631\{0.592\} & 1.02\{1.1\} & K_1^\prime\rho  & 0.628\{0.777\} & 1.21\{1.69\}\\
 K(1460)\omega  & 0.629\{0.786\} & 1.02\{1.46\} & Kf_2 & 0.61\{0.615\} & 1.17\{1.34\}\\ 
 \multicolumn{3}{c|}{\cdots}     & \multicolumn{3}{c}{\cdots}  \\ \hline
 \text{Total} & 100\{100\} & 162\{186\} & \text{Total} & 100\{100\} & 192\{218\}\\
\hline
 \hline
\end{array}\]
\end{table*}

\subsection{High spin kaons with $J^{P}=4^{\pm}$}
\begin{figure}[htbp]
\centering%
\includegraphics[scale=1]{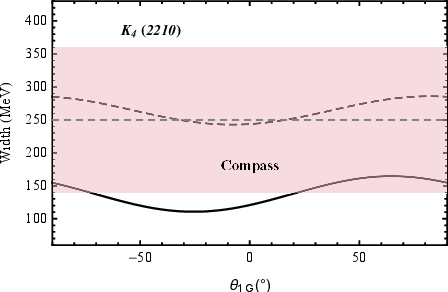}
\caption{{\color{black}The dependence of the total decay width on the mixing angle $\theta$ for the $K_4(2210)$ state, along with corresponding experimental data \cite{COMPASS:2025wkw} (pink band) for comparison with our theoretical calculation. The solid line corresponds to the result obtained using the experimental mass, while the dashed line denotes that obtained with the theoretical mass derived from the MGI model. }}
\label{K42210}
\end{figure}

For $J^P=4^{-}$ states, the $K_4(2210)$ has been observed by the COMPASS Collaboration in the $\pi K_2^*$ (D wave) channel recently \cite{COMPASS:2025wkw}. 
The strong decays of $K_4(2210)$ and $K_4^\prime(1G)$ are presented in Table \ref{K41G}. Assuming the mixing angle $\theta_{1G}=-41.8^ \circ$, the calculated total decay width is slightly smaller than the experimental value \cite{COMPASS:2025wkw}. 
Notably, $K a_2$, $\rho K_1$, $\rho K^*$, $K f_2$, $K \rho_3(1690)$, and $\pi K^*$ are the main decay channels of $K_4(2210)$ as a $1G$ state. The decay channel $K a_2$ is the most prominent, accounting for a branching ratio of 0.23. In addition, the branching ratio of $\pi K_2^*$ \purple{reaches $2\%$,} which is consistent with the experimental result reported by \purple{COMPASS Collaboration} \cite{COMPASS:2025wkw}.   
The $\theta_{1G}$ dependence of the total decay width of $K_4(2210)$ is depicted in Fig.~\ref{K42210}. When the mixing angle $\theta_{1G}$ is in the range of $[-90^\circ, -71^\circ]$ and $[21^\circ, 90^\circ]$, the total decay width of $K_4(2210)$ overlaps with experimental value \cite{COMPASS:2025wkw}. When the theoretical mass of $K_4(1G)$ is used, the derived width is in good agreement with the experimental value throughout the entire range of mixing angle values as shown in Fig.~\ref{K42210}.

For $K_4(1G)$ and $K_4^\prime(1G)$, we use the same experimental value of mass as input for the calculation.
In the context of a $1G$ state, we predict the mass of $K_4^\prime(1G)$ to be 2306 MeV. 
The $K_4^\prime(1G)$ dominantly decays into $\pi K_3^*(1780)$.  The $\pi K_4^*(2045)$, $\pi K_2^*$, $\rho K^*$, $K \rho$, and $K^* b_1$ are the important decay modes. These channels are conducive to their identification in experimental research.

The PDG only lists the $K_4(2500)$ state. This state has been observed in the decay process $K^-p \to \Lambda \bar{p}p$ by the Geneva-Lausanne spectrometer, with a mass of $M=2490 \pm 20$  MeV and a width of $\Gamma \sim 150$ MeV, as reported in 1981 \cite{Cleland:1980ya}. 
Fortunately, the COMPASS Collaboration recently observed the $K_4$ ground state, with a mass and width being  $2210 \pm 40 ^{+80}_{-30}$ MeV and $250 \pm 70 ^{+50}_{-70}$ MeV, respectively \cite{COMPASS:2025wkw}. 
With our current theoretical understanding, we conclude that $K_4(2210)$ appears to be a viable candidate for the ground state of the $K_4$ family, whereas $K_4(2500)$ is considered as a $2G$ state.

\begin{figure}[htbp]
\centering%
\includegraphics[scale=1]{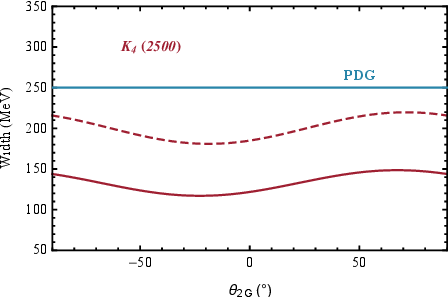}
\caption{{\color{black}The $\theta$ dependence of the total decay width of $K_4(2500)$ state, along with the corresponding experimental data \cite{ParticleDataGroup:2024cfk} (represented as a turquoise blue line) for comparison with our theoretical calculation. The solid line corresponds to the result obtained using the experimental mass, while the dashed line denotes that obtained with the theoretical mass derived from the MGI model.}}
\label{K421G4}
\end{figure}

The decay widths of $K_4(2500)$ and $K_4^\prime(2G)$ are calculated using the same experimental mass.
The $K_4(2500)$ has been identified as the most suitable candidate for the $2G$ state. 
Its counterpart, $K_4^\prime(2G)$, has also undergone investigation and their OZI-allowed two-body strong decay behaviors are further elaborated in Table \ref{K42G}. 
For $K_4(2500)$ and $K_4^\prime(2G)$, the widths calculated using the theoretical mass as input are larger than those derived with the experimental mass.
The  important decay channels of $K_4(2500)$ include $Ka_2$, $K a_2(1700)$, $K \rho_3(1690)$, $K \omega_3(1945)$, $\rho K_1$, $K f_2$, and $\rho K^*(1410)$, with corresponding branching ratios of 0.11, 0.090, 0.077, 0.058, 0.052, 0.040, and 0.038, respectively. When the theoretical mass is used for the initial state, the total width of $K_4(2500)$ is 187 MeV. Meanwhile, the most important decay modes for $K_4(2500)$ is $K \rho_3(1990)$.
The dependence of the total decay width of $K_4(2500)$ on the mixing angle $\theta_{2G}$ is illustrated in Fig. \ref{K421G4}. The calculated total decay width is lower than the experimental value \cite{ParticleDataGroup:2024cfk}, for which only the central value is currently available experimentally. We look forward to the release of more experimental data on $K_4(2500)$ in the future. 

For the $K_4^\prime(2G)$ state, no experimental observation has been reported thus far. Our theoretical analysis suggests that the predominant decay channels involve the channels $\pi K_3^*(2D)$, $\pi K_3^*(1780)$, $\pi K_2^*$, $\pi K_4^*(2045)$, and $K^* b_1$. Additional information can be found in Table \ref{K42G}.

\begin{table*}[htbp]
\renewcommand{\arraystretch}{1.3}
\centering
\color{black}
\caption{The decay information of $K_4(2210)(1G)$ and $K_4^\prime(1G)$ with the mixing angle $\theta_{1G}=-41.8^\circ$. 
The unit of the width is MeV. \label{K41G}}
\[\begin{array}{ccc|ccccc}
\hline
\hline
\multicolumn{3}{c|}{K_4(2210), \Gamma_{\text{Exp}}=250\pm70^{+50}_{-70}~$\cite{COMPASS:2025wkw}$}&\multicolumn{3}{c}{K_4^\prime(1G)}\\
\hline
\text{Channel}  & \text{Br(\%)}    & \text{Value}    & \text{Channel}    & \text{Br(\%)} & \text{Value} \\\midrule[1pt]
 Ka_2             &22.5\{11.4\}     &26.3\{29.2\}     & \pi K_3^*(1780)   &26.9\{17.5\}     &43.2\{47.7\}\\
 \rho K_1         &9.74\{7.55\}     &11.4\{19.3\}     & \pi K_4^*(2045)   &16.9\{18.9\}     &27.1\{51.6\}  \\
 \rho K^*         &8.2\{5.78\}      &9.56\{14.8\}     & \pi K_2^*         &13.7\{7.75\}     &22\{21.1\} \\
 Kf_2             &7.95\{3.76\}     &9.27\{9.62\}     & \rho K^*          &5.93\{5.42\}     &9.52\{14.8\} \\
K\rho_3(1690)     &7.22\{21.4\}     &8.42\{54.9\}     & K\rho             & 4.47\{4.07\}    &7.18\{11.1\}\\
\pi K^*           &6.82\{4.6\}      &7.95\{11.8\}     & K^*b_1            & 3.7\{8.33\}     & 5.95\{22.7\} \\
 K\omega_3(1670)  &5.36\{8.18\}     & 6.25\{21\}     & K^*h_1            & 3.19\{3.8\}     & 5.11\{10.4\} \\
 \pi K^*(1410)    &3.91\{3.83\}     & 4.56\{9.81\}    & Ka_1              & 2.7\{3.18\}     & 4.33\{8.67\} \\
K\rho             &3.82\{1.62\}     &4.46\{4.15\}     & \pi K^*(1410)     &2.56\{1.25\}     &4.1\{3.4\}  \\
\pi K_1           &3.71\{2.48\}     & 4.33\{6.34\}    & \rho K_1          & 2.55\{3.28\}    & 4.1\{8.94\}  \\
 \omega K_1       & 3.05\{2.47\}    & 3.56\{6.33\}    & \pi K^*           & 2.5\{1.29\}     & 4.01\{3.51\}  \\
 \omega K^*       & 2.67\{1.89\}    & 3.11\{4.84\}    & \eta K_2^*        & 2.3\{1.98\}     & 3.69\{5.4\}  \\
 K^*a_1           & 2.61\{4.8\}     & 3.04\{12.3\}    & \omega K^*        & 1.93\{1.77\}    & 3.1\{4.83\}  \\
 \pi K_2^*        & 1.85\{1.89\}    & 2.16\{4.86\}    & K\omega           & 1.46\{1.33\}    & 2.34\{3.63\}  \\
\eta K_2^*       & 1.01\{0.819\}   & 1.18\{2.1\}      & \pi K_1           & 0.946\{0.814\}  &  1.52\{2.22\}  \\
\pi K_1^\prime    & 0.968\{0.743\}  & 1.13\{1.9\}     & Ka_2              & 0.878\{1.13\}   & 1.41\{3.08\}  \\
\eta K^*          & 0.913\{0.721\}  & 1.06\{1.85\}    & \omega K_1        & 0.793\{1.06\}   & 1.27\{2.88\}  \\
\rho K_2^*        & \cdots\{3.34\}  & \cdots\{8.55\}  &K^*a_2            &\cdots\{2.61\}    &\cdots\{7.12\}\\ 
\rho K_1^\prime   & \cdots\{1.43\}   & \cdots\{3.68\} &K^*f_2            & \cdots\{1.72\}   & \cdots\{4.69\} \\
K^*a_2           &\cdots\{1.08\}     &\cdots\{2.77\}  &\rho K_1^\prime   &\cdots\{1.25\}    & \cdots\{3.41\}  \\ 
\omega K_2^*     &\cdots\{0.982\}    &\cdots\{2.52\}  &\pi K_2^*(1980)   & \cdots\{1.12\}   & \cdots\{3.06\}\\
\pi K_2 (1770)    &\cdots\{0.953\}   &\cdots\{2.44\}  &\rho K_2^*        &\cdots\{1.07\}    &\cdots\{2.91\}\\ 
K^*f_1          & \cdots\{0.783\}  & \cdots\{2.01\}   &Kf_1            & \cdots\{0.762\} & \cdots\{2.08\}  \\
K^*f_2          & \cdots\{0.71\}   & \cdots\{1.82\}   &Kf_2              & \cdots\{0.493\}  & \cdots\{1.34\} \\ 
K^*b_1            &\cdots\{0.479\} & \cdots\{1.23\}   &\rho K^*(1410)    &\cdots\{0.395\}   &\cdots\{1.08\}\\
K^*h_1          & \cdots\{0.479\}  & \cdots\{1.23\}   &\pi K_2(1770)     & \cdots\{0.387\}  & \cdots\{1.05\} \\ 
K\rho(1450)      &\cdots\{0.446\}  &\cdots\{1.14\}    &\omega K_1^\prime &\cdots\{0.377\}   & \cdots\{1.03\} \\  
 \pi K_1^\prime(2P)&\cdots\{0.443\}   &\cdots\{1.14\} &\multicolumn{3}{c}{\cdots}\\     
\omega K_1^\prime &\cdots\{0.436\}  &\cdots\{1.12\}  && &\\ 
 \rho K^*(1410)   &\cdots\{0.433\}   &\cdots\{1.11\}    & & &\\ 
   \pi K^*(1680)    &\cdots\{0.433\}  &\cdots\{1.11\}    & & &\\ 
\multicolumn{3}{c|}{\cdots}                           &   & &\\ \hline
 \text{Total}    & \text{100\{100\}} & \text{117\{256\}} & \text{Total} & \text{100\{100\}} & \text{161\{273\}} \\
 \hline
\hline
\end{array}\]
\end{table*}

\begin{table*}[htbp]
\renewcommand{\arraystretch}{1.3}
\centering
\color{black}
\caption{The decay information of $K_4(2500)(2G)$ and the  $K_4^\prime(2G)$ with the mixing angle $\theta_{2G}=-41.8^\circ$. 
The unit of the width is MeV. \label{K42G}}
\[\begin{array}{ccc|ccccc}
\hline
 \hline
   \multicolumn{3}{c|}{K_4(2500), \Gamma_{\text{Exp}}\approx 250}&\multicolumn{3}{c}{K_4^\prime(2G)}
  \\
\hline
 \text{Channel} & \text{Br(\%)} & \text{Value} & \text{Channel} & \text{Br(\%)} & \text{Value} \\
\midrule[1pt]
Ka_2            & 10.8\{7.93\}   & 13.1\{14.8\}     & \pi K_3^*(2D)   & 13.2\{8.51\} & 19.1\{18.2\}  \\
 Ka_2(1700)     & 8.87\{5.49\}   & 10.8\{10.3\}     & \pi K_3^*(1780) & 12.6\{8.92\} & 18.3\{19.1\}\\
K\rho_3(1690)   & 7.6\{8.09\}    & 9.24\{15.1\}     & \pi K_2^*       & 9.32\{6.36\} & 13.5\{13.6\} \\
K\omega_3(1945) &5.76\{6.37\}    & 7\{11.9\}       & \pi K_4^*(2045) & 9.21\{6.85\} & 13.4\{14.7\} \\
\rho K_1        &5.48\{4.2\}     & 6.67\{7.84\}     & K^*b_1          & 3.77\{4.08\} & 5.47\{8.75\} \\
 Kf_2           & 3.96\{2.8\}    & 4.81\{5.23\}     & \pi K_2^*(1980) & 3.46\{1.93\} & 5.02\{4.14\} \\
 \rho K^*(1410) &3.81\{3.15\}    & 4.63\{5.88\}     & \rho K^*(1410)  & 3.21\{2.8\}  & 4.66\{5.99\}\\
 K\omega_3(1670)& 3.21\{3.16\}   & 3.9\{5.9\}       & Ka_1(1640)      & 3.01\{2.86\} & 4.37\{6.12\}\\
 K\rho          & 2.98\{1.96\}   & 3.62\{3.66\}     & \pi K^*         & 2.86\{1.92\} & 4.15\{4.1\}\\
 Kf_2(2P)       & 2.96\{1.34\}   & 3.59\{2.51\}     & K\rho (1450)    & 2.46\{2.14\} & 3.58\{4.59\}\\
 K^*a_1         & 2.77\{2.62\}   & 3.37\{4.9\}      & \rho K^*        & 2.15\{1.78\} & 3.12\{3.81\}  \\
 \rho K^*       & 2.61\{2.1\}    & 3.17\{3.93\}     & \rho K_1        & 2.13\{1.8\}  & 3.09\{3.86\} \\
 \rho K_2^*     & 2.44\{2.76\}   & 2.96\{5.15\}     & K^*h_1          & 2.06\{1.93\} & 2.99\{4.13\}  \\
\pi K_1^\prime(2P)& 2.05\{1.25\} & 2.49\{2.33\}     & K^*a_2          & 1.79\{2.17\} & 2.6\{4.64\}  \\
 \pi K^*        & 2\{1.83\}     & 2.44\{3.42\}     & \pi K^*(3S)     & 1.68\{1.24\} & 2.45\{2.65\}  \\
 \omega K_1     & 1.78\{1.37\}   & 2.16\{2.57\}     & \eta K_2^*      & 1.56\{1.27\} & 2.26\{2.72\}  \\
 \pi K_2^*      & 1.57\{1.19\}   & 1.91\{2.23\}     & \pi K_2^*(3P)   & 1.51\{1.76\} & 2.19\{3.77\}  \\
 \pi K_1        & 1.32\{1.17\}   & 1.6\{2.18\}      & \rho K_2^*      & 1.2\{1.37\}  & 1.75\{2.93\}  \\
 \pi K^*(1410)  & 1.24\{\cdots\}  & 1.51\{\cdots\}  & K\rho           & 1.17\{1.16\} & 1.69\{2.48\} \\
\omega K^*(1410) & 1.23\{1.02\}  & 1.5\{1.92\}      & \omega K^*(1410)& 1.04\{0.91\} & 1.51\{1.95\}\\
 K^*b_1          & 1.21\{0.842\} & 1.48\{1.57\}     & \rho K_1^\prime & 0.999\{1\}  & 1.45\{2.14\} \\
 K^*\pi (1300)   & 1.07\{1.18\}  & 1.3\{2.2\}       & Ka_2            & 0.998\{0.877\} & 1.45\{1.88\} \\
 \rho K_1^\prime & 1.04\{1.18\}  & 1.27\{2.21\}     & K\omega (1420)  & 0.986\{0.82\} & 1.43\{1.76\} \\
 Ka_4(2040)      & 1.04\{2.11\}  & 1.27\{3.95\}     & K^*f_2          & 0.963\{1.04\} & 1.4\{2.24\} \\
 K^*a_2          & 0.999\{1.18\} & 1.21\{2.21\}     & K^*\pi(1300)    & 0.946\{0.533\}& 1.37\{1.14\} \\
 K\omega         & 0.991\{0.653\}& 1.2\{1.22\}      & Ka_2(1700)      & 0.74\{0.889\} & 1.07\{1.9\} \\
 \rho K(1460)    & 0.932\{\cdots\}& 1.13\{\cdots\}  & \rho K(1460)    & 0.719\{0.866\}& 1.04\{1.86\}\\
K\rho_3(1990)   & \cdots\{13.6\} &\cdots\{25.4\}    & \omega K^*      & 0.703\{0.583\}& 1.02\{1.25\} \\
 \pi K^*(3S)     & 0.894\{1.15\} & 1.09\{2.15\}     & \omega K_1      & 0.691\{0.589\}& 1\{1.26\} \\
K^*\rho (1450)  & \cdots\{1.05\} & \cdots\{1.95\}   & K^*\rho(1450)   & \cdots\{0.897\} & \cdots\{1.92\}\\
 Ka_2(1950)      & \cdots\{1.02\} & \cdots\{1.91\}  &Kf_2^\prime(1525)& \cdots\{0.559\} &\cdots\{1.2\}\\
 Kf_4(2050)      & \cdots\{1.01\} & \cdots\{1.89\}  &Kf_1(2P)         & \cdots\{0.59\} & \cdots\{1.26\} \\ 
\pi K_2^*(1980) & \cdots\{0.934\} & \cdots\{1.75\}  &\eta K_3^*(1780) &\cdots\{0.603\} & \cdots\{1.29\}\\
\omega K_2^*    &\cdots\{0.869\} &\cdots\{1.62\}    & Kf_2(2P)        & \cdots\{0.684\} & \cdots\{1.47\} \\
 \omega K^*      & 0.854\{0.689\}& 1.04\{1.29\}     & \multicolumn{3}{c}{\cdots}\\
 \pi K_3^*(1780)  & \cdots\{0.679\} & \cdots\{1.27\} & &&\\
 \pi K_2^*(1F)   & \cdots\{0.587\} &\cdots\{1.1\}   &&&\\
 K^*f_2          & \cdots\{0.586\} & \cdots\{1.09\} & &&\\
   K^*f_1          & \cdots\{0.55\} & \cdots\{1.03\}  & &&\\
 \multicolumn{3}{c|}{\cdots}                        & &&\\ \hline
 \text{Total} & \text{100\{100\}} & \text{122\{187\}} & \text{Total} & \text{100\{100\}} & \text{145\{214\}} \\
 \hline
 \hline
\end{array}\]
\end{table*}

For $J^P=4^{+}$ states, $K_4^*(2045)$ was initially observed in the $K^+d\to K^+X$ reaction in 1977. 
It has a mass of $2115 \pm 46$ MeV and a width of $300 \pm 200$ MeV. This particle decays into $K\pi$, $K^*\pi$, $K\rho$, $K^*\rho$, $K^*(1420)\pi$, and $Kf_2$, with a $4^+$ spin-parity assignment being strongly supported \cite{Carmony:1977ju}. 
In 1981, using the LASS spectrometer at SLAC, researchers gathered high-statistics data from the reaction $K^-p \to K^-\pi^+n$. The partial waves displayed evidence of a $J^P=4^+$ $K^*$ resonance. The resonance's mass and width were determined to be $2070^{+100}_{-40}$ MeV and $240^{+500}_{-100}$ MeV, respectively \cite{Aston:1981th}.
Subsequently, in 1982, the $K_4^*(2045)$ resonance was observed again by SPEC and HBC \cite{Birmingham-CERN-Glasgow-MichiganState-Paris:1982tev, Cleland:1982td}. 
In 1986, new measurement of the mass and width of the $4^+$ $K^*(2045)$ \red{[$K^{*}(2045)$ is equal to $K_4^{*}(2045)$]} resonance \purple{was conducted.} This measurement indicates that the mass and width of $K_4^*(2045)$ are $2062\pm 14\pm 13$ MeV and $221 \pm 48 \pm27$ MeV, respectively \cite{Aston:1986rm}. In the same year, Torres \red{\textit{et al}}. investigated the inclusive final states of $pN\to K^+K^-K^+K^- X$ at $400$ GeV$/c$, and they presented evidence for the decays $K^{*0}(2045)\to \phi K^\pm \pi^\mp$ and $K^{*0}(2045)\to \phi K^{*0}(890)$ \red{[which is named $K^{*}(2060)$ in their work]}
\cite{Torres:1985yi}.

The mass of the $K_4^*(2045)$ we obtained is 2075 MeV, which is in approximately consistent with experimental measurements $2060\pm5^{+11}_{-3}$ MeV  \cite{COMPASS:2025wkw} and $2048^{+8}_{-9}$ MeV \cite{ParticleDataGroup:2024cfk}.
$K_4^*(2045)$ is considered as a strong candidate for the $1F$ state and the OZI-allowed two-body strong decay width is 111 MeV. Since the theoretical mass is close to the experimental mass, the calculated widths also show little difference.
The $\rho K^*$ channel is the dominant decay mode, with a branching ratio of 0.34. Other decay channels, including $\omega K^*$, $\pi K^*$, $K \rho$, $\pi K$, $\pi K_2^*$, and $Kb_1$, also make significant contributions to the total width of $K_4^*(2045)$.
In particular, the final state $\pi K$ has been experimentally identified with a measured partial width of $0.099 \pm 0.012$ MeV \cite{ParticleDataGroup:2024cfk}, a value that closely matches our computed result. Further details are provided in Table \ref{K4star}.

\begin{table}[htbp]
\renewcommand{\arraystretch}{1.3}
\centering
\color{black}
\caption{The decay information of  $K_4^*(2045)(1^3F_4)$. 
The unit of the width is MeV.  \label{K4star}   }
\[\begin{array}{cccc}
\hline
 \hline
 \text{Channel} &\text{Br\%}
 & \text{Exp.}~$\cite{ParticleDataGroup:2024cfk}$ &\text{Value} \\
 \hline
 \rho K^*       &33.5\{31.9\}   &         &37.1\{39.4\}  \\
 \omega K^*     &11\{10.5\}    &         &12.2\{13\}  \\
 \pi K^*        &8.44\{8.28\}   &         &9.35\{10.2\} \\
K\rho           &7.71\{7.66\}   &         &8.53\{9.47\} \\
\pi K           &7.61\{7.25\}   &9.9\pm1.2\%&8.43\{8.97\} \\
 \pi K_2^*      &4.65\{4.93\}   &         &5.15\{6.1\} \\
 Kb_1           &3.91\{4.27\}   &         &4.33\{5.29\} \\
 \pi K_1        &3.02\{3.14\}   &         &3.34\{3.89\} \\
K\omega         &2.51\{2.49\}   &         &2.78\{3.08\} \\
\pi K_1^\prime  &2.49\{2.56\}   &         &2.76\{3.17\} \\
 Ka_2           &2.43\{2.89\}   &         &2.69\{3.57\} \\
 \pi K^*(1410)  &2.09\{2.31\}   &         &2.31\{2.86\} \\
 Kh_1           &1.83\{1.92\}   &         &2.03\{2.38\}\\
 \eta K^*       &1.6\{1.65\}    &         &1.77\{2.04\} \\
 Kf_2           &1.37\{1.55\}   &         &1.51\{1.92\} \\
 Ka_1           &1.19\{1.29\}   &         &1.32\{1.6\} \\
 K^*\phi        &1.07\{1.3\}    &1.4\pm0.7\%  &{\color{black}1.02\{1.61\} }\\
\pi K(1460)     &0.914\{1.05\}  &         &1.01\{1.3\} \\ 
\multicolumn{4}{c}{\cdots}  \\ \hline
\text{Total} & 100 \{100\} &199^{+27}_{-19} &  111\{124\}  \\
 \hline
 \hline
\end{array}\]
\end{table}

\subsection{High spin kaons with $J^{P}=5^{\pm}$}

For $J^P=5^-$ state, $K_5^*(2380)$ was initially identified in the $K^- p \rightarrow K^- \pi^+ n$ reaction in 1986, exhibiting a mass and width of $2382\pm14\pm19$ MeV and $178\pm37\pm32$ MeV, respectively \cite{Aston:1986rm}. 
In 1988, a study of $K^-\pi^+$ scattering within $K^- p \rightarrow K^- \pi^+ n$ reaction at a center-of-mass energy of 11 GeV/$c^2$ yielded a branching ratio for $K\pi$ of $6.1 \pm 1.2 \%$ \cite{Aston:1987ir}, comparable to the present results. $K_5^*(2380)$ mainly decays into $K^* a_2$, $\rho K_2^*$, $\rho K^*$, $K^* f_2$, $\omega K_2^*$, $K b_1$, and $K a_2$, with other typical decay branching ratios presented in Table \ref{K5star}. The mass of $K_5^*(1G)$ obtained by the MGI model is 2308 Mev, which is lower than the experimental mass of $2382 \pm 24$ MeV \cite{ParticleDataGroup:2024cfk}. Therefore, the total width calculated using the theoretical mass is also smaller than that obtained from experimental measurements as input. Further details of the OZI-allowed two-body strong decays of $K_5^*(2380)$ are presented in Table \ref{K5star}.

As $1H$ states in $J^P=5^+$, neither $K_5(1H)$ nor $K_5^\prime(1H)$ \purple{has been} observed. We aim to predict their masses and OZI-allowed two-body strong decay behaviors. 
The calculations indicate that the masses of both $K_5(1H)$ and $K_5^\prime(1H)$ are approximately 2509 MeV. 
The decay information for the $K_5(1H)$ and $K_5^\prime(1H)$ is summarized in Table \ref{K51H}. 
The total width of the $K_5(1H)$ meson is calculated to be 276 MeV. 
The decay channels and their corresponding branching ratios include: $K \rho_3(0.14)$, $K a_4(2040) (0.094)$, $\rho K_2^* (0.052)$, $\pi K^*(1410) (0.050)$, $K \omega_3(1670) (0.048)$, and $K^*a_1 (0.043)$, which are accessible for investigation by the \purple{COMPASS}, LHCb, and BESIII Collaborations in future experiments. Concurrently, our \purple{prediction} for the total width of the $K_5^\prime(1H)$ meson is 278 MeV. 
Both $K_5(1H)$ and $K_5^\prime(1H)$ states are broad resonances with widths of approximately 280 MeV, making experimental exploration more demanding. Present calculations indicate that the $\pi K_4^*(2045)$, $\pi K_3^*(1780)$, $\pi K_5^*(2380)$, $K^* b_1$, and $K^* a_2$ channels are particularly significant for the $K_5^\prime(1H)$.

\begin{table}[htbp]
\renewcommand{\arraystretch}{1.1}
\centering
\caption{The decay information of $K_5(1^1H_5)$ and $K_5^\prime(1^3H_5)$ with the mixing angle $\theta_{1H}=-42.4^\circ$. 
The unit of the width is MeV.  \label{K51H} }
\[\begin{array}{ccc|ccccc}
\hline
 \hline
   \multicolumn{3}{c|}{K_5(1H)}&\multicolumn{3}{c}{K_5^\prime(1H)}
  \\
\hline
 \text{Channel} & \text{Br(\%)} & \text{Value} & \text{Channel} & \text{Br(\%)} & \text{Value} \\
\midrule[1pt]
 K\rho_3(1690)     & 14.2           & 39.3      & \pi K_4^*(2045)     & 14.7      & 40.9 \\
 K a_4(2040)       & 9.41           & 26        & \pi K_3^*(1780)     &  8.51     & 23.7 \\
 \rho K_2^*        & 5.23           & 14.4      & \pi K_5^*(2380)     & 7.48      & 20.8 \\
 \pi K^*(1410)     & 4.97           & 13.7      & K^*b_1              & 5.53      & 15.4 \\
 K\omega_3(1670)   & 4.83           & 13.3      & K^*a_2              & 5.47      & 15.2 \\
 K^*a_1            & 4.25           & 11.7      & \rho K^*            &3.52       & 9.8 \\
 K a_2             &3.92            & 10.8      & \rho K_1            &3.16       & 8.79 \\
 \rho K_1          & 3.54           & 9.77      & K a_1               & 3.08      & 8.56 \\
 \rho K^*          & 3.53           & 9.73      & \pi K_2^*           & 2.7       & 7.5 \\
 \pi K^*           & 2.61           & 7.2       &K \rho               & 2.49      & 6.93 \\
 K\phi_3(1850)     & 2.48           & 6.85      & K^*f_2              & 2.26      & 6.28 \\
 \pi K_2^*         & 2.27           & 6.25      & \pi K_2^*(1980)     & 2.25      & 6.25 \\
 K^*b_1            & 2.25           & 6.2       & \pi K_3^*(2D)       & 2.23      & 6.2 \\
 \rho K^*(1410)    & 2.19           & 6.03      & \rho K_2^*          & 2.17      & 6.02 \\
 K^*a_2            & 2.16           & 5.97      & \rho K^*(1410)      & 2.16      & 6 \\
 \omega K_2^*      & 1.7            & 4.68      & \eta K_3^*(1780)    & 2.09      & 5.82 \\
 \pi K_1           & 1.56           & 4.3       & \rho K_1^\prime     & 1.96      & 5.45 \\
 \rho K_1^\prime   & 1.55           & 4.29      & K^*h_1              & 1.95      & 5.42 \\
 \pi K_2(1770)     & 1.46           & 4.02      & K \rho _2(1D)       & 1.53      & 4.25 \\
 K f_4(2050)       & 1.39           & 3.85      & K a_2               & 1.39      & 3.87 \\
 K f_2             & 1.23           & 3.4       & \omega K^*          & 1.16      & 3.21 \\
K^*h_1             & 1.21           & 3.34      & \omega K_1          & 1.03      & 2.86 \\
\omega K_1         & 1.18           & 3.25      &K \rho(1450)         & 0.842     & 2.34 \\
\omega K^*         & 1.16           & 3.19      & K^*a_1              & 0.836     & 2.32 \\
\rho  K(1460)      & 1.13           & 3.13      & K \omega            & 0.819     & 2.28 \\
 K a_2(1700)       & 1.12           & 3.08      & K f^\prime_2(1525)  & 0.813     & 2.26 \\
K^*f_1             & 0.936          & 2.58      & K a_1(1640)         & 0.802     & 2.23 \\
K^*f_2             & 0.882          & 2.43      & \eta K_2^*          & 0.713     & 1.98 \\
 \pi K^*(1680)     & 0.805          & 2.22      & \omega K_2^*        & 0.699     & 1.94 \\
 \phi K_1          & 0.707          & 1.95      & \omega K^*(1410)    & 0.685     & 1.9 \\
\omega K^*(1410)   & 0.694          &1.92       &K \rho(1700)         & 0.681     & 1.89 \\
 \pi K_3^*(1780)   & 0.659          &1.82       &\omega K_1^\prime    & 0.632     & 1.76 \\
 \eta K_3^*(1780)  & 0.656          &1.81       & K f_1               & 0.626     & 1.74 \\
\pi K_1^\prime     & 0.647          & 1.79      & K^* \pi(1300)       & 0.619     & 1.72 \\
\pi K_0^*(1430)    & 0.614          &1.69       & K^*\phi             & 0.584     & 1.62 \\
K^*\phi            & 0.588          & 1.62      & K f_2               & 0.571     & 1.59 \\
\pi K_1^\prime(2P) & 0.56           &1.54       & K a_0               & 0.548     & 1.52 \\
\omega K_1^\prime  & 0.506          &1.4        & \pi K^*(1410)       & 0.522     & 1.45 \\
 \pi K_4(2210)     & 0.482          & 1.33      & \pi K_2(1770)       & 0.516     & 1.44 \\
 \eta K_2^*        & 0.448          & 1.24      & K\omega_2(1D)       & 0.509     & 1.42 \\
\eta K^*(1410)     & 0.408          &1.13       & K f_0(1500)         & 0.499     & 1.39 \\
 K \rho            & 0.386          & 1.06      &K \phi               & 0.46      & 1.28 \\
 \eta K^*          & 0.371          & 1.02      & \pi K_1             & 0.415     & 1.15 \\
 K f_2(2P)         & 0.364          & 1         & K \omega(1420)      & 0.387     & 1.07 \\
   \multicolumn{3}{c|}{\cdots}   & K\rho_3(1690) & 0.381  & 1.06 \\
                      &  &       &K \omega(1650) & 0.38   & 1.06 \\
                      &  &        &\multicolumn{3}{c}{\cdots}\\       
\hline
 \text{Total} & 100 & 275 & \text{Total} & 100 & 278 \\
 \hline
 \hline
\end{array}\]
\end{table}

\begin{table}[htbp]
\renewcommand{\arraystretch}{1.2}
\centering
\color{black}
\caption{The decay information of  $K_5^*(1G)(2380)$. 
The unit of the width is MeV. The branching ratio of $\pi K$ is   $0.061\pm0.012$ \cite{ParticleDataGroup:2024cfk}.  \label{K5star}}
\[\begin{array}{cccc}
\hline
 \hline
 \text{Channel}   & \text{Br(\%)}                 & \text{Value} \\ \midrule[1pt]
 K^*a_2           & 14.5\{10.3\}              & 46.2\{17.4\}  \\
 \rho K_2^*       & 14.3\{11.8\}              & 45.3\{19.9\}  \\
 \rho K^*         & 6.63\{10.9\}              & 21.1\{18.4\}  \\
 K^*f_2           & 6.43\{6.54\}              & 20.4\{11\}  \\
 \omega K_2^*     & 4.54\{3.49\}              & 14.4\{5.9\}  \\
 Kb_1             & 3.66\{4.51\}              & 11.6\{7.62\}  \\
 Ka_2             & 3.4\{3.79\}               & 10.8\{6.4\}  \\
 \pi K_2^*        & 3.31\{4.2\}               & 10.5\{7.1\}  \\
 \pi K^*(1410)    & 2.82\{3.35\}              & 8.96\{5.65\} \\
\rho K^*(1410)    & 2.66\{1.58\}              & 8.44\{2.67\}  \\
 \pi K^*          & 2.58\{3.77\}              & 8.19\{6.37\}  \\
 K\rho            & 2.54\{3.6\}               & 8.09\{6.09\}  \\
 \pi K_1          & 2.29\{2.84\}              & 7.27\{4.79\}  \\
 \omega K^*       & 2.2\{3.61\}               & 6.98\{6.09\}  \\
\pi K_3^*(1780)   & 1.98\{1.62\}              & 6.27\{2.73\}  \\
 \pi K            & 1.76\{2.76\}              & 5.6\{4.66\}  \\
 Kh_1             & 1.75\{2.26\}              & 5.55\{3.82\} \\
\pi K_1^\prime    & 1.69\{2.37\}              & 5.36\{4.01\} \\
 \pi K(1460)      & 1.54\{1.69\}              & 4.89\{2.85\}  \\
 Kf_2             & 1.42\{1.67\}              & 4.52\{2.82\}  \\
 Ka_1             & 1.24\{1.54\}              & 3.93\{2.61\}  \\
 K\pi(1300)       & 0.977\{0.963\}            & 3.1\{1.63\}  \\
 \rho K_1         & 0.944\{0.714\}            & 3\{1.21\} \\
 K^*b_1           & 0.915\{\cdots\}            & 2.91\{\cdots\}  \\
 K\rho_3(1690)    & 0.861\{\cdots\}            & 2.73\{\cdots\}  \\
 K\omega          & 0.835\{1.18\}             & 2.65\{1.99\}  \\
 \omega K^*(1410) & 0.826\{\cdots\}             & 2.63\{\cdots\}  \\
 K\pi_2           & 0.61\{\cdots\}             & 1.94\{\cdots\}  \\
 K\eta            & 0.565\{0.816\}            & 1.8\{1.38\}  \\
 K^*h_1           & 0.565\{\cdots\}             & 1.79\{\cdots\}  \\
 \pi K_1(1790)    & 0.552\{\cdots\}            & 1.75\{\cdots\}  \\
\pi K_4^*(2045)   &0.512\{\cdots\}            & 1.63\{\cdots\}  \\
 Kf_1             & 0.488\{\cdots\}            & 1.55\{\cdots\}  \\
 \eta K_2^*       & 0.472\{\cdots\}            & 1.5\{\cdots\}  \\
 K^*\phi          & 0.462\{\cdots\}            & 1.47\{\cdots\}  \\
\pi K_2(1820)     & 0.406\{\cdots\}             & 1.29\{\cdots\}  \\
K\omega_3(1670)   &0.399\{\cdots\}             & 1.27\{\cdots\}  \\
 K\eta_2          & 0.395\{\cdots\}             & 1.25\{\cdots\}  \\
 K\rho(1450)      & 0.347\{\cdots\}            & 1.1\{\cdots\}  \\
 K\eta(1295)      & 0.337\{\cdots\}            & 1.07\{\cdots\} \\ 
 \multicolumn{3}{c}{\cdots}\\ \hline
\text{Total} & 100 \{ 100 \}              & 318 \{ 169 \}  \\ 
\hline
 \hline
\end{array}\]
\end{table}

\section{Summary}\label{4}

Inspired by the recent observations of the $K_3$ and $K_4$ states, 
we have conducted a comprehensive investigation of high-spin kaons, examining them within the frameworks of the Regge trajectory, the MGI model, and the $^3P_0$ model. In our analysis, we utilize meson spatial space wave functions, derived from the MGI model, to compute the OZI-allowed two-body strong decay widths.

Our results indicate that the $K_3$ and $K_4$ states, recently detected by the COMPASS Collaboration, are promising candidates for the $K_3^\prime(2120)(1F)$ and $K_4(2210)(1G)$ kaons. The decay modes of $K_3^\prime(2120)(1F)$ and  $K_4(2210)(1G)$ we calculated are in agreement with experimental observations.

The partner of $K_3^\prime(2120)(1F)$, $K_3(1F)$, mainly decays into $K a_2$ and $\rho K^*$. For future experimental searches for the $K_3(1F)$ state, researchers can begin by investigating $K3\pi$ final states. The channels $\pi K_3^*(1780)$, $\pi K_4^*(2045)$, and $\pi K_2^*$ will be crucial for testing our proposed assignments and studying the $K_4^\prime(1G)$ state.
 
Furthermore, the predicted widths for the $K_3(1F)$, $K_3^\prime(2F)$, $K_4^\prime(1G)$, $K_4^\prime(2G)$, $K_5(1H)$, and  $K_5^\prime(1H)$ resonances are approximately  
250 MeV, 190 MeV, 160 MeV, 145 MeV, 280 MeV, and 280 MeV, respectively, and their OZI-allowed two-body strong decays are investigated. 
This work also further verifies the applicability of the MGI model and the $^3P_0$ model.

We anticipate that upcoming experimental studies will play a pivotal role in elucidating the nature of the newly observed kaons and in validating or further exploring these theoretical predictions.

\begin{acknowledgments}
This work is supported by the National Natural Science Foundation of China under Grants  No.~12235018, No.~11975165, No.~11965016, and No.~12247101, and by the Natural Science Foundation of Qinghai Province under Grant No. 2022-ZJ-939Q, the Fundamental Research Funds for the Central Universities (Grant No. lzujbky-2024-jdzx06).
\end{acknowledgments}

\bibliographystyle{apsrev4-1}
\bibliography{hepref}
\end{document}